\documentclass[proof]{WileyASNA-v1}
\usepackage{comment}
\usepackage{pdfpages}

\articletype{Article}%

\received{20 July 2020}
\revised{22 September 2020}
\accepted{22 September 2020}

\begin{document}
% we only have measurements between 1998 and 2018, so Twenty year feels more accurate than the quarter century
\title{The HD 217107 Planetary System: Twenty Years of Radial Velocity Measurements}

\author[1]{Mark R. Giovinazzi}
\author[1]{Cullen H. Blake}
\author[2]{Jason D. Eastman}
\author[3]{Jason Wright}
\author[4]{Nate McCrady}
\author[5]{Rob Wittenmyer}
\author[2]{John A. Johnson}
\author[7]{Peter Plavchan}
\author[1]{David H. Sliski}
\author[2]{Maurice L. Wilson}
\author[6]{Samson A. Johnson}
\author[5]{Jonathan Horner}
\author[8]{Stephen R. Kane} 
%\author[2]{Andrew Szentgyorgyi}
\author[4]{Audrey Houghton}
\author[2]{Juliana Garc\'{i}a-Mej\'{i}a}
\author[9]{Joseph P. Glaser}

\authormark{Giovinazzi \textsc{et al.}}

\address[1]{\orgdiv{Department of Physics and Astronomy}, \orgname{University of Pennsylvania}, \orgaddress{\state{Pennsylvania}, \country{USA}}}

\address[2]{\orgdiv{Center for Astrophysics \textbar \ Harvard \& Smithsonian}, \orgaddress{\state{Massachusetts}, \country{USA}}}

\address[3]{\orgdiv{Department of Astronomy / Center for Exoplanets and Habitable Worlds / Penn State Extraterrestrial Intelligence Center},\orgname{The Pennsylvania State University} \orgaddress{\state{Pennsylvania}, \country{USA}}}

\address[4]{\orgdiv{Department of Physics and Astronomy},\orgname{The University of Montana} \orgaddress{\state{Montana}, \country{USA}}}

\address[5]{\orgdiv{Centre for Astrophysics},\orgname{University of Southern Queensland} \orgaddress{\state{Toowoomba}, \country{Australia}}}

\address[6]{\orgdiv{Department of Astronomy},\orgname{The Ohio State University} \orgaddress{\state{Ohio}, \country{USA}}}

%\address[7]{\orgdiv{Department of Physics and Astronomy},\orgname{University of Utah} \orgaddress{\state{Utah}, \country{USA}}}

\address[7]{\orgdiv{Department of Physics and Astronomy},\orgname{George Mason University} \orgaddress{\state{Virginia}, \country{USA}}}

\address[8]{\orgdiv{Department of Earth and Planetary Sciences},\orgname{University of California at Riverside} \orgaddress{\state{California}, \country{USA}}}

\address[9]{\orgdiv{Department of Physics}, \orgname{Drexel University} \orgaddress{\state{Pennsylvania}, \country{USA}}}

\corres{Mark Giovinazzi \\
\email{markgio@sas.upenn.edu}}

\presentaddress{Department of Physics and Astronomy, University of Pennsylvania, Philadelphia, PA 19103}

\abstract{The hot Jupiter HD 217107~b was one of the first exoplanets detected using the radial velocity (RV) method, originally reported in the literature in 1999.  Today, precise RV measurements of this system span more than 20 years, and there is clear evidence for a longer-period companion, HD 217107~c. Interestingly, both the short-period planet ($P_\mathrm{b}\sim7.13~$d) and long-period planet ($P_\mathrm{c}\sim5059$~d) have significantly eccentric orbits ($e_\mathrm{b}\sim0.13$ and $e_\mathrm{c}\sim0.40$). We present 42 additional RV measurements of this system obtained with the MINERVA telescope array and carry out a joint analysis with previously published RV measurements from four different facilities. We confirm and refine the previously reported orbit of the long-period companion. HD 217107~b is one of a relatively small number of hot Jupiters with an eccentric orbit, opening up the possibility of detecting precession of the planetary orbit due to General Relativistic effects and perturbations from other planets in the system. In this case, the argument of periastron, $\omega$, is predicted to change at the level of $\sim$0.8$^\circ$~century$^{-1}$. Despite the long time baseline of our observations and the high quality of the RV measurements, we are only able to constrain the precession to be $\dot{\omega}<65.9^\circ$~century$^{-1}$. We discuss the limitations of detecting the subtle effects of precession in exoplanet orbits using RV data.}


\maketitle

\section{Introduction}\label{sec1}

The announcement in 1995 of the first exoplanet detected using the Radial Velocity (RV) technique, 51 Peg b (\citealt{mayor1995}), marked the beginning of a period of rapid growth in our knowledge of planets orbiting stars other than our Sun. In the five years that followed the discovery of 51 Peg b, a handful of additional planets were found - most of which were more massive than Jupiter, but moved on surprisingly small orbits, with periods of just a few days to a few weeks \footnote{see \url{http://exoplanets.org} for chronology of discovery}. These unexpected short period planets soon became known as ``hot Jupiters". HD 217107~b was one of the first hot Jupiters discovered, initially reported in \citet{fischer1999} as having a period of $P=7.12$~d and a minimum mass of $m$~$\sin{i}=1.27$~M$_{\rm{J}}$ based on 21 RV measurements with a typical precision of 6~m~s$^{-1}$, more than sufficient to reveal the star's large RV semi-amplitude of $K\sim140$~m~s$^{-1}$. Later observations by \citet{vogt2000, naef2001,vogt2005, wittenmyer2007, wright2009} revealed the presence of a long-term RV trend that emerged as the clear signal of an outer planet with a period of more than 10 years. \citet{feng2015} presented a joint analysis of the existing RV data and determined a complete orbit for HD 217107~c with a period of $P=5189\pm21$~d and $m$~$\sin{i}=4.153\pm$0.017~M$_{\rm{J}}$. 

Today, RV observations of some of the first exoplanets discovered span more than two decades. These long time baselines enable the search for not only outer companions with orbits similar to that of Jupiter in our solar system, but also subtle effects that may cause the orbits of the inner planets to change over time. For example, in General Relativity (GR), the orientation of the orbit of a planet will evolve in time as the argument of periastron precesses. The orbital precession of the planet Mercury in our own solar system is well known and measured to be 56'' year$^{-1}$, of which 43'' century$^{-1}$ is due to GR effects. As discussed by, among others, \citet{jordi2001}, \citet{kane2012}, and \citet{Jord_n_2008}, the gravitational quadrupole moment of the star and an outer-perturbing planet can also cause precession, but those effects are generally much smaller for the case of an exoplanet. Figure 1 depicts the phenomenon. These precession effects are difficult to measure in exoplanet systems, but may provide interesting tests of GR and models of stellar structure, as well as clues to the existence of undetected outer companions.

\begin{figure}[h!]
	\centerline{\includegraphics[width=78mm,height=60mm]{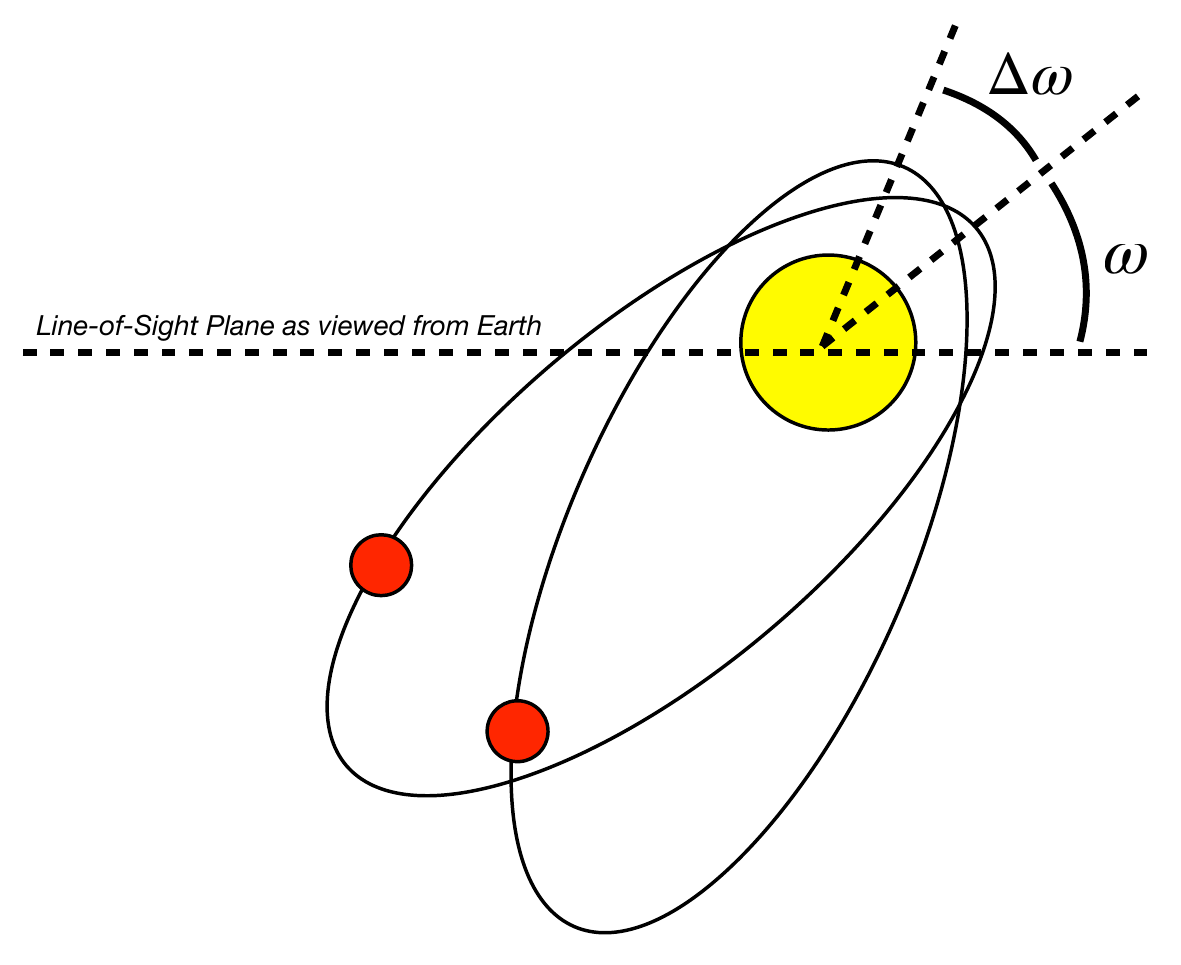}}
	\caption{Argument of periastron, $\omega$, is defined as the angle between the line-of-sight plane as viewed from Earth and the planet's point of periastron. In GR, planetary orbits are not necessarily closed and $\omega$ can change over time. Over some period of time, $\Delta t$, the orbit itself will rotate around the star by some angle $\Delta \omega$; this yields $\dot{\omega}={\Delta \omega}/{\Delta t}$.}
	\label{precession_diagram}
\end{figure}

We present a joint analysis of 377 RV observations of the HD 217107 system spanning 20.3 years. The 42 new measurements presented here, obtained with the MINERVA telescope array (\citealt{swift2015, wilson2019}), extend the total published observational time baseline by almost five years. We confirm the existence of the outer companion, HD 217107~c, and refine the orbital solutions for both components. We model the orbital precession of HD 217107~b and constrain it to be $\dot{\omega}<65.9 ^{\circ} \rm{century}^{-1}$ (95$\%$ confidence), an upper limit two orders of magnitude above the expected level of precession due to GR effects.

\section{Stellar Parameters}

HD 217107 is a main-sequence star 20.3~pc away ($\pi=49.817\pm0.057$~mas; \citealt{gaiadr2}) and is similar in mass to our Sun. We used the SED fitting capabilities within EXOFASTV2  (\citealt{eastman2017}) to estimate the stellar properties of HD 217107 using the known distance and broad-band photometry. We estimated the stellar mass to be 1.09$^{+0.065}_{-0.071}$~$\mathrm{M}_\odot$, the radius to be 1.140$^{+0.039}_{-0.036}$~$\mathrm{R}_\odot$, the effective temperature to be 5670$^{+110}_{-100}$~K, and the metallicity to be $\rm{[Fe/H]}=0.415^{+0.067}_{-0.072}$. The estimated surface gravity is $\log{g}=4.362^{+0.041}_{-0.048}$, indicating that HD 217107 is a somewhat evolved yellow subgiant star (\citealt{stassun2017}; \citealt{Cubillos_2011}; \citealt{wittenmyer2007}).

\section{Data Sets} We analyzed published RV data sets from the HIRES, Hamilton, CORALIE, and Robert G. Tull Coud\'e spectrometers, in addition to the new MINERVA RVs presented here for the first time. We use 128 Keck-{\sc i} RV measurements published in \citet{feng2015} and break the full data set into two parts corresponding to before and after the HIRES CCD upgrade that took place in 2004 (denoted as HIRES and HIRES2, respectively, throughout this analysis). We use 121 measurements obtained with the Hamilton spectrometer at Lick Observatory as published in \citet{wright2009}, 23 RV measurements obtained with the Robert G. Tull Coud\'e at McDonald Observatory as published in \citet{wittenmyer2007}, and 63 RV measurements obtained with the CORALIE instrument at Observatoire de Haute-Provence as reported in \citet{naef2001}. The typical reported RV uncertainty on these measurements is between 2.5 and 7.5 m s$^{-1}$. In all cases we converted the reported times of observations, JD$_{\rm{UTC}}$ or BJD$_{\rm{UTC}}$, to BJD$_{\rm{TDB}}$ following \citet{eastman2010} in order to facilitate direct comparisons over a time span of two decades.

The MINERVA array is a set of four 0.7~m telescopes at the Fred Lawrence Whipple Observatory in Arizona. MINERVA observed HD 217107 between May 2016 and November 2018 and the derived RV measurements are reported in Table~1. The MINERVA observatory, which is described in more detail in \citet{swift2015} and \citet{wilson2019}, has been demonstrated to achieve an RV precision of $\sigma_{\mathrm{RV}}\sim2$~m~s$^{-1}$ on bright RV standard stars. The four telescopes are fiber-coupled to a KiwiSpec echelle spectrometer with a resolution of $R\sim84{,}000$ and a spectral coverage of approximately 500-600~nm. Four spectra, one from each telescope, are recorded simultaneously and the four derived RVs are combined. The spectrometer is stabilized in terms of temperature and pressure and uses an I$_{2}$ absorption cell for calibration of the wavelength solution and instrumental profile. 

\begin{center}
\begin{table}[h!]%
\centering
\caption{MINERVA RV Measurements\label{MINERVA_measurements}}%
\tabcolsep=0pt%
\begin{tabular*}{20pc}{@{\extracolsep\fill}lcc@{\extracolsep\fill}}
\toprule
\textbf{Time (BJD$_{\rm{TDB}}$)} & \textbf{RV (m s$^{-1}$)}  & \textbf{$\sigma_{\rm{RV}}$ (m s$^{-1}$)}  \\
\midrule
2457531.95052  &   148.95  &   5.67\\
2457532.94528  &    18.36  &   5.31\\
2457533.94540  &   -64.68  &   5.39\\
2457534.94600  &   -74.95  &   5.32\\
2457746.67498  &    44.62  &   5.75\\
2457749.65332  &    -6.48  &   5.60\\
2458013.79343  &    -2.62  &   6.39\\
2458014.79785  &   111.80  &   5.32\\
2458014.82302  &   117.47  &   5.34\\
2458018.66054  &   -86.83  &   6.06\\
2458018.89143  &   -86.63  &   7.34\\
2458019.73547  &   -81.30  &   6.42\\
2458020.88991  &   -30.12  &   7.87\\
2458023.71030  &   142.42  &   6.69\\
2458024.73673  &    -4.62  &   7.20\\
2458025.73526  &   -79.90  &   7.06\\
2458026.74815  &   -97.34  &   6.10\\
2458030.85046  &   127.45  &   6.49\\
2458033.67697  &   -91.08  &   5.95\\
2458037.67320  &   173.83  &   6.14\\
2458040.72114  &   -91.73  &   6.15\\
2458040.85717  &   -91.52  &   6.06\\
2458043.67616  &   139.18  &   6.42\\
2458045.72767  &    58.24  &   5.65\\
2458046.74262  &   -55.11  &  10.88\\
2458050.68397  &   132.03  &   6.41\\
2458052.70612  &    68.54  &   6.27\\
2458053.72147  &   -39.66  &   5.78\\
2458055.66905  &   -62.48  &   5.98\\
2458056.69235  &    15.60  &   5.81\\
2458060.67300  &   -41.74  &   8.16\\
2458063.66764  &    -5.03  &   5.95\\
2458081.62238  &    28.64  &   5.67\\
2458083.62211  &   -84.46  &   5.76\\
2458429.62483  &   162.07  &   5.55\\
2458430.62032  &    40.33  &   5.44\\
2458431.62414  &   -68.00  &   5.36\\
2458432.62643  &  -110.29  &   5.43\\
2458433.62326  &   -70.04  &   5.47\\
2458434.61911  &    12.17  &   6.02\\
2458435.59711  &   128.74  &   5.77\\
2458439.62649  &  -103.08  &   5.50\\

\bottomrule
\end{tabular*}
\begin{tablenotes}
%\item Source: Example for table source text.
%\item[$\dagger$] Example for a first table footnote.
%\item[$\ddagger$] Example for a second table footnote.
\end{tablenotes}
\end{table}
\end{center}

\section{Orbital Analysis}
We carried out an analysis of the combined RV data set using RadVel: The Radial Velocity Fitting Toolkit (\citealt{fulton2018}). We fit for five Keplerian orbital parameters for each planet: Period ($P$), eccentricity ($e$), argument of periastron ($\omega$), time of inferior conjunction ($T_{\rm{conj}}$), and RV semi-amplitude ($K$). At the same time, we fit for an RV offset ($\gamma$) between each facility (one offset for each of the two parts of the HIRES data set), along with an additional RV ``jitter'' term ($\sigma$) for each facility, and a long-term RV trend ($\dot{\gamma}$). We use the RadVel fitting basis that corresponds to fitting the Keplerian orbital parameters directly.

RadVel is a Python package that fits for the Keplerian orbital parameters through maximum \textit{a posteriori} optimization and a Markov Chain Monte Carlo (MCMC) approach to sampling the posterior distributions of the parameters and estimating confidence intervals. We place uniform priors, corresponding to physical bounds, or wide regions around previously published parameters, on all parameters except eccentricity (see Table 2). We do not specifically exclude scenarios where the periastron distance of planet b is comparable to the radius of the host star. Biases resulting from different priors when estimating small values of eccentricity from noisy data have been discussed in the literature (\citealt{shen2008, gregory2010,nadia2011}). We adopt the Gaussian prior used in \citet{tuomi2012}, which is an approximation of the observed eccentricity distribution from known exoplanets. Given that the eccentricities $e_\mathrm{b}$ and $e_\mathrm{c}$ are relatively large, and that we have a considerable number of RV measurements, we expect that the choice of eccentricity prior has little impact on the estimated orbital solution. We find consistent results in fits with the Gaussian prior and a simple uniform prior on eccentricity. 

\begin{center}
\begin{table}[t]%
\centering
\caption{Prior distributions used as input to Radvel MCMC analysis. A prior of [$p_1$, $p_2$] denotes a uniform bound between $p_1$ and $p_2$. A prior of N[$\mu$, $\sigma$] denotes a Gaussian prior centered at $\mu$ with standard deviation $\sigma$.}%
\label{prior_dists}
\tabcolsep=0pt%
\begin{tabular*}{20pc}{@{\extracolsep\fill}ccc@{\extracolsep\fill}}
\toprule
Parameter & Prior & Units \\
\midrule
\centering
$P_\mathrm{b}$ & [7.126746, 7.126946] & days \\
$T_\mathrm{conj_b}$ &  [2452889.51658, 2452896.64342] & days \\
$e_\mathrm{b}$ & N[0, 0.3]$\cup$[0, 0.99] & -- \\
$\omega_\mathrm{b}$ & [-$\pi$, $\pi$] & rad\\
$K_\mathrm{b}$ & [130.30, 150.30] & m~s$^{-1}$ \\
$P_\mathrm{c}$ &  [4650, 5450] & d \\
$T_\mathrm{conj_c}$ & [2452411.38, 2457461.38] & days\\
$e_\mathrm{c}$ & N[0, 0.3]$\cup$[0, 0.99] & -- \\
$\omega_\mathrm{c}$ & [-$\pi$, $\pi$] & rad \\
$K_\mathrm{c}$ & [0, 500] & m~s$^{-1}$ \\
$\sigma_\mathrm{HIRES}$ & [0, 25] & m~s$^{-1}$ \\
$\sigma_\mathrm{HIRES2}$ & [0, 25] & m~s$^{-1}$ \\ 
$\sigma_\mathrm{CORALIE}$ & [0, 25] & m~s$^{-1}$ \\
$\sigma_\mathrm{Hamilton}$ & [0, 25] & m~s$^{-1}$ \\
$\sigma_\mathrm{CES}$ & [0, 25] & m~s$^{-1}$ \\
$\sigma_\mathrm{MINERVA}$ & [0, 25]  & m~s$^{-1}$ \\
\bottomrule
\end{tabular*}
\end{table}
\end{center}

The results of the RadVel fits are given in Table~3, and the best-fit orbital solution and residuals are shown in Figure~2. In Figure~3 we show a subset of the posterior distributions from the MCMC analysis, demonstrating that the orbital parameters are well-measured. The orbital parameters are consistent with the most recent literature values from \citet{feng2015}, though our analysis prefers a slightly shorter period for the outer planet, $P_\mathrm{c}=5059^{+52}_{-49}$~d, approximately 131~d less than the period reported in \citet{feng2015}. All other orbital parameters are consistent with \citet{feng2015} within the reported uncertainties. We note that an overall RV trend, $\dot{\gamma}\sim0.0025$~m~s$^{-1}$, is detected at at the 2.5$\sigma$ level. This may be evidence for an additional companion with a period longer than 20 years. However, we note that there is strong covariance between $P_\mathrm{c}$, $K_\mathrm{c}$, and $\dot{\gamma}$, meaning that the long-term RV trend may be an artefact of the relatively sparse time coverage of the long HD 217107~c orbit and the RV offsets between facilities (HIRES2 to MINERVA - see Figures 2 and 3). We considered a three-planet model and found a solution with a third companion with low significance at P$_\mathrm{d}$=12283$\pm440$~d, K$_\mathrm{d}$=13$\pm$13 m s$^{-1}$ and $e_\mathrm{d}$=0.21$\pm0.17$. We note that this period is longer than the duration of our RV data set, so any estimate of the orbital parameters of this potential planet d are highly degenerate with the RV offset, $\gamma_{\rm{MINERVA}}$, for the MINERVA measurements relative to the HIRES, CORALIE, and CES measurements.

\begin{figure}
	\centerline{\includegraphics[width=0.5\textwidth]{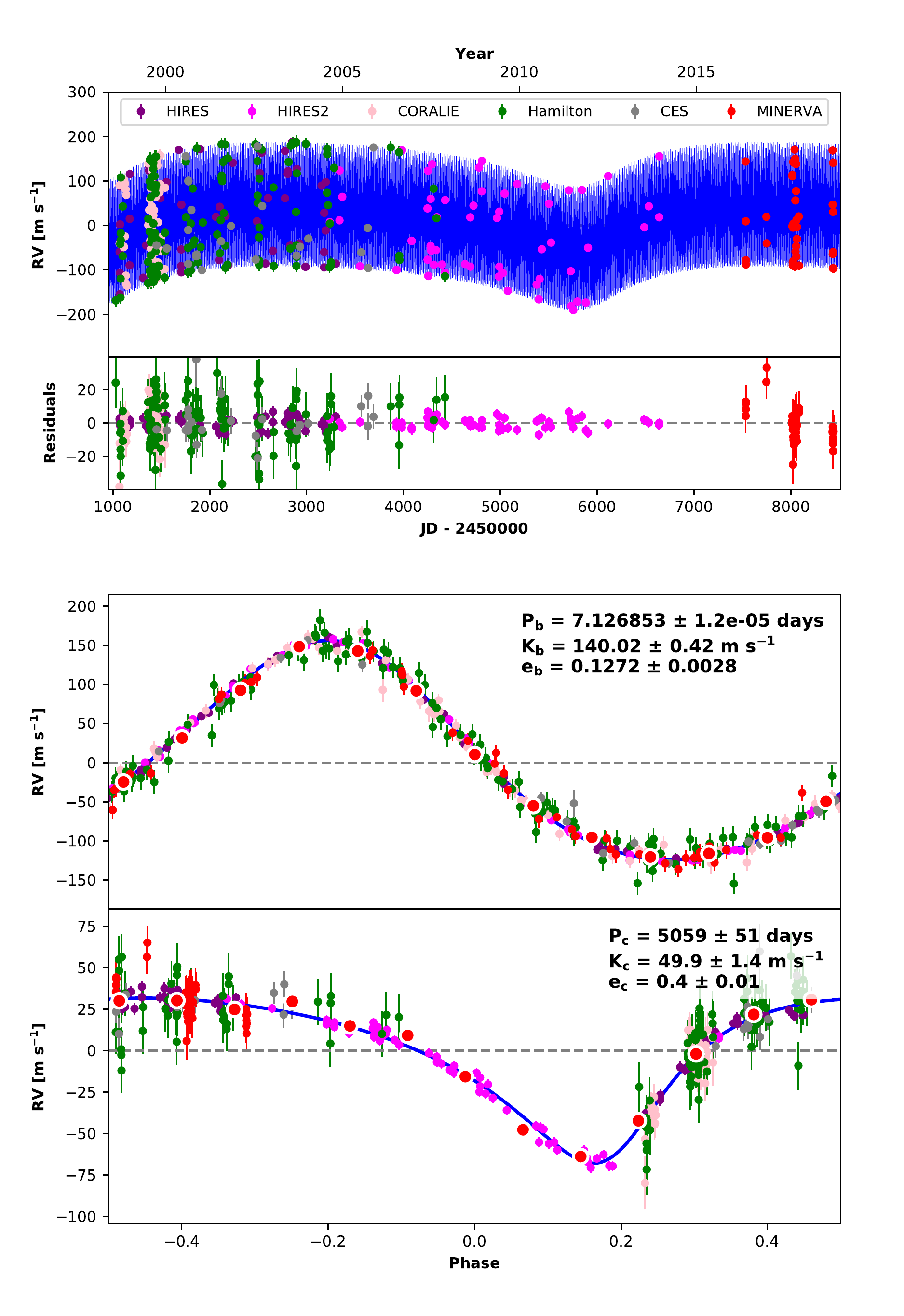}}
	\caption{ HD 217107 orbital solution. \textbf{Top:} The full RV time series, along with the best-fit model and RV residuals. Different facilities are denoted by point color, and HIRES2 corresponds to the post-upgrade HIRES measurements. \textbf{Middle and Bottom:} Phase-folded RV curves for HD 217107~b and HD 217107~c. Large red points are averages in bins of phase.} \label{RV_multipanel}
\end{figure}

\begin{figure}
	\centerline{\includegraphics[width=0.5\textwidth]{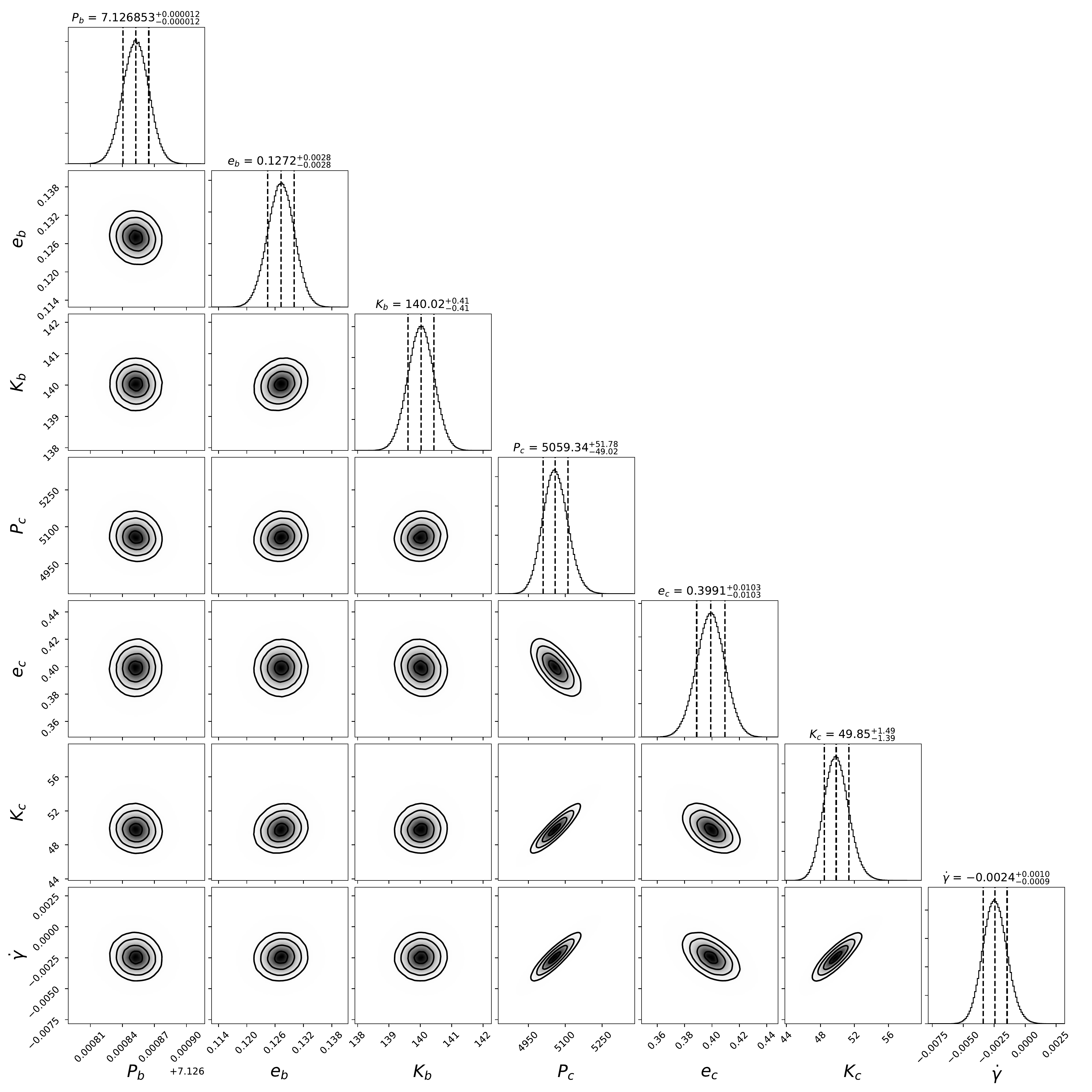}}
	\caption{Select sample of posterior distributions from our RadVel MCMC analysis.} \label{RV_multipanel}
\end{figure}

\begin{comment}
\begin{figure*}[hbpt]
	\centerline{\includegraphics[width=0.8\textwidth]{HD217107_rv_multipanel.pdf}}
	\caption{This is how the multipanel would look on a whole page. I thought it would be helpful at first, but I think it's only more distracting. Leaving here for reference, but will likely delete this one...} \label{RV_multipanel}
\end{figure*}
\end{comment}

%\includepdf[]{HD217107_rv_multipanel.pdf}
\begin{center}
\begin{table}[t]%
\centering
\caption{HD 217107 System Best-fit Orbital and Derived Parameters. The Maximum Likelihood values are given for each instrument's RV ``jitter'' term since we define those to be >0, and the posteriors are therefore non-Gaussian. We fit for an RV offset between each instrument (breaking the HIRES data into two instruments) but do not include those values in this table. \label{best-fits}}%
\tabcolsep=0pt%
\begin{tabular*}{20pc}{@{\extracolsep\fill}ccc@{\extracolsep\fill}}
\toprule
Parameter & Credible Interval & Units \\
\midrule
\centering
$P_\mathrm{b}$ & $7.126853\pm1.2e-05$ & days \\
$T_\mathrm{conj_b}$ & $2452893.0794^{+0.0067}_{-0.0066}$ & days \\
$e_\mathrm{b}$ & $0.1272\pm0.0028$ & --\\
$\omega_\mathrm{b}$ & $0.419\pm0.023$ & rad \\
$K_\mathrm{b}$ & $140.02\pm0.42$ & m~s$^{-1}$ \\
$a_\mathrm{b}$ & $0.0746^{+0.0015}_{-0.0016}$ & AU \\ 
$m_\mathrm{b}~\sin~i_\mathrm{b}$ & $1.394^{+0.057}_{-0.059}$ & M$_\mathrm{J}$ \\
$P_\mathrm{c}$ & $5059.34^{+52.78}_{-49.02}$ & days \\
$T_\mathrm{conj_c}$ & $2454950^{+24}_{-25}$ & days \\
$e_\mathrm{c}$ & $0.3991\pm0.0103$ & --\\
$\omega_\mathrm{c}$ & $3.572\pm0.031$ & rad \\
$K_\mathrm{c}$ & $49.85^{+1.49}_{-1.39}$ & m~s$^{-1}$ \\
$a_\mathrm{c}$ & $5.94\pm0.13$ & AU \\ 
$m_\mathrm{c}~\sin~i_\mathrm{c}$ & $4.09^{+0.23}_{-0.224}$ & M$_\mathrm{J}$ \\
$\dot{\gamma}$ & $-0.00245^{+0.00099}_{-0.00095}$ & m~s$^{-1}$~$\mathrm{day}^{-1}$ \\
\midrule
\midrule
Parameter & Maximum Likelihood & Units \\
\midrule
$\sigma_\mathrm{HIRES}$ & 0.0 & m~s$^{-1}$ \\
$\sigma_\mathrm{HIRES2}$ & 0.3 & m~s$^{-1}$ \\ 
$\sigma_\mathrm{CORALIE}$ & 0.0 & m~s$^{-1}$ \\
$\sigma_\mathrm{Hamilton}$ & 13.0 & m~s$^{-1}$ \\
$\sigma_\mathrm{CES}$ & 4.6 & m~s$^{-1}$ \\
$\sigma_\mathrm{MINERVA}$ & 8.6  & m~s$^{-1}$ \\
\bottomrule
\end{tabular*}
\begin{tablenotes}
%\item Source: Example for table source text.
%\item[$\dagger$] Example for a first table footnote.
%\item[$\ddagger$] Example for a second table footnote.
\end{tablenotes}
\end{table}
\end{center}

% another possible way..
\begin{comment}
\begin{tabular*}{20pc}{m{2pc}|m{5pc}m{8pc}m{4pc}}%{@{\fill}lcc@{\fill}lcc@{\fill}}
\toprule
&Parameter & Credible Interval & Units \\
\midrule
\parbox[t]{-10mm}{\multirow{7}{*}{\rotatebox[origin=c]{90}{HD 217107b}}} \\
 & ~~~$P_$ & $7.126853\pm1.2e-05$ & days \\
 & ~$T_\mathrm{conj}$ & $2452893.0793^{+0.0067}_{-0.0066}$ & days \\
 & $e$ & $0.1272\pm0.0028$ & --\\
 & $\omega$ & $0.419\pm0.023$ & rad \\
 & $K$ & $140.02\pm0.41$ & m~s^{-1} \\
 & $a$ & $0.07503^{+0.00096}_{-0.00098}$ & AU \\ 
 &$m~\sin~i$ & $1.41^{+0.036}_{-0.037}$ & M_J \\
 \midrule
\parbox[t]{-10mm}{\multirow{7}{*}{\rotatebox[origin=c]{90}{HD 217107c}}}
 & $P_\mathrm{c}$ & $5058^{+52}_{-49}$ & days \\
 & $T_\mathrm{conj_c}$ & $2454951^{+24}_{-25}$ & days \\
 & $e_\mathrm{c}$ & $0.3996\pm0.0103$ & --\\
 & $\omega_\mathrm{c}$ & $3.572\pm0.031$ & rad \\
 & $K_\mathrm{c}$ & $49.83^{+1.50}_{-1.38}$ & m~s^{-1} \\
 & $a_\mathrm{c}$ & $5.975\pm0.087$ & AU \\ 
 & $m_\mathrm{c}~\sin~i_\mathrm{c}$ & $4.13^{+0.18}_{-0.17}$ & M_J \\

\end{tabular*}
\end{comment}

\section{Orbital Precession}
In GR, the orbit of a planet is not necessarily closed as it is, in the absence of other perturbers, in Newtonian gravity. This results in a slow evolution of the argument of periastron of an orbit. Famously, this effect was measured in the orbit of Mercury in our own solar system as one of the first direct tests of GR. The size of the GR effect is given as
\begin{eqnarray}
\dot{\omega}_\mathrm{GR}=\frac{7.78}{\left(1-e^2\right)}\frac{M_*}{\mathrm{M}_\odot}\left(\frac{a}{0.05\mathrm{AU}}\frac{P}{\mathrm{day}}\right)^{-1}\left[\circ/\mathrm{century}\right]
\end{eqnarray}
In Table~4 we report the known exoplanets expected to exhibit GR precession at the level of $10^{\circ}$ century$^{-1}$ or more. 

Other effects can also lead to orbital precession, including the quadrupole moment of the star and the influence of massive, outer companions. Attempts at measuring the precession of an exoplanet orbit have been made (see, for example, \citealt{Iorio2011,Figueira_2016, noriharu2020}). \citet{Figueira_2016} used high-precision RV data in an attempt to detect a predicted precession in HD 80606~b of $\dot{\omega}\sim0.06^\circ$~century$^{-1}$, but doing so with only RV measurements proved difficult. \citet{Figueira_2016} were only able to place an upper limit of $\dot{\omega}<2.7\pm3.1^\circ$~century$^{-1}$. Transit observations, with their sharp features that can be precisely timed, may provide a better opportunity for measuring orbital precession in exoplanet systems. For example, \cite{Blanchet_2019} proposed that, given future transit data from the James Webb Space Telescope (JWST), $\dot{\omega}$ for HD 80606~b could be easily measured. 

Given the long time baseline and high RV precision of our observations, we attempted to measure $\dot{\omega}$ for the HD 210107~b orbit. We developed a code to directly incorporate $\dot{\omega}$ into the two-planet Keplerian orbit fits by including a time-dependent term of the form
$\omega(t)=\omega_{0}+\dot{\omega}\cdot t$. Starting from the best-fit orbital parameters determined with RadVel, we refit the RV data with the $\dot{\omega}$ term using maximum \textit{a posteriori} optimization and an MCMC approach to explore the posterior distributions of the parameters. In this case, we used the differential evolution MCMC code within EXOFASTV2 (\citealt{eastman2017}) to carry out the MCMC analysis. With our data, we find that we are only able to rule out very large values of $\dot{\omega}$ and with $95\%$ confidence constrain $\dot{\omega}<65.9^\circ$~century$^{-1}$. This is a factor of $\sim$80 greater than the predicted precession for the orbit of HD 217107~b, which is 0.81$^\circ$~century$^{-1}$. 

To test the reliability of our code, and explore the quantity and quality of RV data that would be required to measure orbital precession in HD 217107~b, we generated synthetic observations at the actual times of the observations of our RV data set and differing levels of Gaussian noise. We included in these synthetic data sets $\dot{\omega}=0.81^\circ$~century$^{-1}$ and attempted to recover the known value of the precession of the periastron. This question was considered by \citet{Jord_n_2008} who estimated that 100 individual measurements having 1~m~s$^{-1}$ precision gathered over 20 years may be sufficient in some cases to detect precession in massive, short-period planets on eccentric orbits. We found that in the case of very low noise $\left(\sigma_{\mathrm{RV}}=0.1~\mathrm{m}~\mathrm{s}^{-1}\right)$, our algorithm successfully recovers the known $\dot{\omega}$ to high accuracy given the 377 simulated measurements over 20 years. However, for realistic measurement precision, these simulations indicated that detecting precession effects in exoplanet orbits with RV data alone is going to be very challenging. For example, we found that $\sigma_{\mathrm{RV}}=0.15$~m~s$^{-1}$ would be required to detect precession of 0.81$^\circ$~century$^{-1}$ at $95\%$ confidence in a data set identical in total duration and cadence to our actual observations. At the level of precision readily achievable with modern instruments, 2~m s$^{-1}$, our simulations indicate that more than 50 years of once-per-week observations would be required to detect the 0.81$^\circ$~century$^{-1}$ precession.

In addition to GR effects, the influence of an outer companion may cause orbital precession of HD 217107~b. In our own solar system, it is the effect of the other planets that are responsible for the vast majority of Mercury's precession. \citet{Jord_n_2008} give an approximation for the effect of perturbing bodies, $\omega_\mathrm{pert}$, to first order in eccentricity and lowest order in the ratio of the semi-major axes of the two planets as follows.
\begin{eqnarray}
\dot{\omega}_{\rm{pert}}\approx29.6\left(\frac{P}{\mathrm{day}}\right)^{-1}\left(\frac{a_\mathrm{b}}{a_\mathrm{c}}\right)^3\left(\frac{M_cM_{\odot}}{M_\oplus M_{*}}\right)[\circ/\mathrm{century}]
\end{eqnarray}
Given that the ratio of the semi-major axes of the two HD~217107 planets is so small ($[a_\mathrm{b}/a_\mathrm{c}]^{3}\sim2\times10^{-6}$), the perturbing effect is expected to be substantially smaller than the GR effect in the orbit of HD 217107~b. Planet c is expected to induce a precession of $0.01^\circ$ century$^{-1}$, roughly 1\% of the GR component of $\dot{\omega}$. Finally, effects related to star-planet tides and the quadrupole moment of the star are also discussed in \citet{Jord_n_2008} and are expected to be very small for HD 217107~b. We note that precession effects, as well as other dynamical effects related to the interaction of planets b and c, may be more readily detectable through Transit Timing Variations (see, for example, \citealt{horner2020,kane2012}). However, we are not aware of the detection of any transits of the host star by planet b, and HD~217107 has not been observed by the Transiting Exoplanet Survey Satellite (\textit{TESS}). HD 217107 was observed as part of the extended \textit{Kepler} mission \textit{K2}. Observations spanning approximately two weeks seemingly show no signs of a transit signal at the period of HD~217107~b. Though the expected amplitude of the transit is large, $\sim$1$\%$, the \textit{K2} light curve exhibits significant systematic effects. If future photometry can confirm that there is no transit signal, we could constrain the inclination of HD~217107~b to be $i<86^{\circ}$, but we are currently unable to rule out the possibility that the system's planets transit.
% dec -2, too close to ecliptic for TESS sadly.

%Owing to the high precision of the orbital elements presented within this paper, we wished to further confine the structure of this system via long-term stability arguments. 
We investigated the long-term dynamical stability of the HD 217107 system within the angular momentum deficit framework (AMD; \citealt{Laskar2017}, \citealt{Petit2017}, \citealt{Petit2018}, \citealt{Glaser2020}), which considers orbital overlap, mean motion resonances (MMR), and Hill instabilities. 
By exploring the inclination parameter space of a co-planar two-planet system, we found that the system is dynamically stable on a timescale equivalent to the lifespan of HD 217107 for inclinations $i>{23.7}^{\circ}$ with an AMD coefficient $\beta^\mathrm{AMD}<$1.

\section{Conclusions}\label{sec5}
We present a joint analysis of RV measurements of the HD~217107 planetary system spanning 20.3 years. In addition to measurements previously reported in the literature, we include 42 new measurements obtained with the MINERVA telescope array for a total of 377 RV measurements. We confirm the existence of a massive, long-period companion on an eccentric orbit (HD 217107~c) and refine its orbital parameters. Given that HD~217107~b has a short-period orbit with a significant eccentricity, we consider the possibility of measuring the precession of the argument of periastron, $\dot{\omega}$, due to GR and other effects. We find that with our current data we are only able to constrain $\dot{\omega}$ to $<65.9^{\circ}$ century$^{-1}$, a level approximately 80 times larger than the expected precession due to GR effects alone. We find that our mean RV precision would need to improve by approximately a factor of 40 to $\sigma_\mathrm{RV}<0.2~\mathrm{m}~\mathrm{s}^{-1}$ in order to expect to detect GR precession in the orbit of HD 217107~b at high significance. We note that there are nine known exoplanet systems for which the inner planet's GR precession is expected to be greater than 10$^{\circ}$ century$^{-1}$ (see Table 4). While our simulations indicate that a data set similar to our HD 217107 data set in terms of precision and extent is not quite sufficient to detect GR precession, modest improvements in overall RV precision could make this possible for the systems in Table 4. 
%\textit{Here is where we will add the small note on stability -- results coming today! (July 15)}
% Joe Glaser Working Here!

%Insert Math range
\begin{center}
\begin{table}[t]%
\centering
\caption{Known exoplanets expected to have $\dot{\omega} >10^\circ~\mathrm{century}^{-1}$. These are planets that have been discovered via Doppler spectroscopy or the transit method. Orbital period and magnitude, in V or \textit{Kepler} bands, are given, along with the estimated $\dot{\omega}$. Even though the predicted precessions are large, some of these targets are faint enough that obtaining precise RV measurements will be challenging. 
 \label{other_precessions}}%
\tabcolsep=0pt%
\begin{tabular*}{20pc}{@{\extracolsep\fill}lccc@{\extracolsep\fill}}
\toprule
Planet & Magnitude & Period [d] & $\dot{\omega}$ [$^\circ$/century] \\
\midrule
\centering
KOI 13 b & 9.96 Kepler & 1.77 & 10.42 \\
HATS 70 b & 12.57 V & 1.89 & 10.43 \\
WASP 114 b & 12.74 V & 1.55 & 11.36 \\ 
GJ 3138 b & 10.98 V & 1.22 & 11.42 \\
HATS 67 b & 13.65 V & 1.61 & 11.52 \\
Kepler 17 b & 14.14 Kepler & 1.49 & 11.72 \\
HATS 52 b & 13.67 V & 1.37 & 13.46 \\
KELT 1 b & 10.70 V & 1.22 & 17.10 \\ 
WASP 19 b & 12.59 V & 0.79 & 27.16 \\
\bottomrule
\end{tabular*}
\end{table}
\end{center}

%\section{Probing for Precession}\label{sec4}

%\textit{might be an awkward use of wording for section title}

%Given our 20 years' worth of high-precision RV observations, 

%\backmatter

\section*{Acknowledgments}
MINERVA is a collaboration among the Harvard-Smithsonian Center for Astrophysics, The Pennsylvania State University, the University of Montana, the University of Southern Queensland, and the University of Pennsylvania. MINERVA is made possible by generous contributions from its collaborating institutions and \fundingAgency{Mt. Cuba Astronomical Foundation}, \fundingAgency{The David \& Lucile Packard Foundation}, the National Aeronautics and Space Administration \fundingAgency{NASA EPSCOR, ExEP, and Nancy Grace Roman programs} (EPSCOR grant NNX13AM97A, Blake is partially supported by a Nancy Grace Roman Fellowship), \fundingAgency{The Australian Research Council} (ARC LIEF grant LE140100050), and the \fundingAgency{National Science Foundation} (NSF grants 1516242 and 1608203 and Graduate Research Fellowships awarded to Giovinazzi and Wilson). Funding for MINERVA data-analysis software development is provided through a subaward under a NASA award NASA MT-13-EPSCoR-0011. This work was partially supported by funding from the Center for Exoplanets and Habitable Worlds, which is supported by the Pennsylvania State University, the Eberly College of Science, and the Pennsylvania Space Grant Consortium. Plavchan is supported in part by NASA Exoplanet Exploration Program, NSF grant AAG 1716202 and George Mason University startup funds. We are grateful to Dr. Gillian Nave and R. Paul Butler for providing FTS measurements of our iodine gas cell. We thank Dr. Matt Holman for answering questions about orbital precession., and Dr. Ben Pope for assistance with \textit{K2} light curves.  Any opinions, findings, and conclusions or recommendations expressed are those of the author and do not necessarily reflect the views of the National Science Foundation. We would like to thank the referee, Dr. Artie Hatzes, for helpful and encouraging feedback on this manuscript.

\subsection*{Author Contributions}

Giovinazzi carried out the RadVel analyses. Blake carried out the analyses and simulations related to constraining $\dot{\omega}$.  Eastman is the MINERVA lead and is responsible for the operation of the MINERVA array.  Sliski assisted with MINERVA operations. Wright provided input on orbit fitting, mechanisms of orbital evolution, and feedback on the manuscript, McCrady is responsible for the optimization of MINERVA observing schedules. Wittenmyer, Wright, J. A. Johnson, and McCrady are founding PIs of the MINERVA project. Wilson assisted in maintaining the MINERVA array as well as developing its operational code and Doppler pipeline. Houghton assisted with remote operations of the MINERVA array. S. A. Johnson assisted in developing MINERVA operational code and maintaining the array. Plavhcan, Kane, and Horner provided feedback on the manuscript. Garc\'{i}a-Mej\'{i}a assisted with the MINERVA spectrograph installation and has provided observatory operational support on site. Glaser carried out the AMD calculations and subsequent simulations to constrain the orbital configuration of HD 217107 b, c, and the theoretical third companion, HD 217107 d.

\subsection{Financial Disclosure}

None reported.

\subsection{Conflict of Interest}

The authors declare no potential conflict of interests.
%\appendix

%\section{Section title of first appendix\label{app1}}

%Use \verb+\begin{verbatim}...\end{verbatim}+ for program codes without math. Use \verb+\begin{alltt}...\end{alltt}+ for program codes with math. Based on the text provided inside the optional argument of \verb+\begin{code}[Psecode|Listing|Box|Code|+\hfill\break \verb+Specification|Procedure|Sourcecode|Program]...+ \verb+\end{code}+ tag corresponding boxed like floats are generated. Also note that \verb+\begin{code}[Code|Listing]...+ \verb+\end{code}+ tag with either Code or Listing text as optional argument text are set with computer modern typewriter font.  All other code environments are set with normal text font. Refer below example:

%\begin{lstlisting}[caption={Descriptive Caption Text},label=DescriptiveLabel]
%for i:=maxint to 0 do
%begin
%{ do nothing }
%end;
%Write('Case insensitive ');
%WritE('Pascal keywords.');
%\end{lstlisting}

%\subsection{Subsection title of first appendix\label{app1.1a}}

%\subsubsection{Subsection title of first appendix\label{app1.1.1a}}

%\noindent\textbf{Unnumbered figure}

%\begin{center}
%\includegraphics[width=7pc,height=8pc,draft]{empty}
%\end{center}

%== Figure 4 ==
%% Example for figure inside appendix

%\nocite{*}% Show all bib entries - both cited and uncited; comment this line to view only cited bib entries;
\bibliography{Wiley-ASNA}%

%\section*{Author Biography}
%\begin{biography}{\includegraphics[]{}}{\textbf{Author Name.} This is sample author biography text this is sample author biography text this is sample author biography text this is sample author biography text this is sample author biography text this is sample author biography text this is sample author biography text this is sample author biography text this is sample author biography text .}
%\end{biography}

\end{document}